\documentclass[final,5p,times,twocolumn]{elsarticle}

\usepackage{amsmath,amssymb,amsfonts}

\usepackage{booktabs}
\usepackage{array}
\usepackage{makecell}

\newcolumntype{C}[1]{>{\centering\arraybackslash}m{#1}}
\usepackage{multirow}
\usepackage{tabularx}
\usepackage{makecell}
\usepackage{etoolbox}
\usepackage{xcolor}
\usepackage{hyphenat}
\usepackage{comment}
\usepackage{ulem}
\usepackage[acronym]{glossaries}
\makeglossaries

\usepackage{pifont}
\usepackage{makecell}
\newcommand{\cmark}{\ding{51}}
\newcommand{\xmark}{\ding{55}}
\usepackage{graphicx}      

\usepackage[table]{xcolor} 

\usepackage{stfloats}      
\usepackage{placeins}      

\usepackage[colorlinks=true,citecolor=red,urlcolor=red]{hyperref}

\usepackage{float}

\usepackage{microtype}

\usepackage{acronym}
\acrodef{IDS}{intrusion detection system}
\acrodef{NIDS}{network intrusion detection system}
\acrodef{HIDS}{host intrusion detection system}
\acrodef{ML}{machine learning}
\acrodef{LLM}{large language model}
\acrodef{RL}{reinforcement learning}
\acrodef{DL}{deep learning}
\acrodef{RAG}{retrieval augmented generation}
\acrodef{VM}{virtual machine}
\acrodef{DoS}{denial of service}
\acrodef{XGBoost}{extreme gradient boosting}

\journal{Computers and Security}

\usepackage{caption}
\begin{document}

\begin{frontmatter}
\title{A Multi-Layer Cloud-IDS Pipeline with LLM and Adaptive Q-Learning Calibration} 
\author{Syed Waqas Ali} 
\affiliation{organization={Department of Computer Software Engineering, University of Engineering and Technology},
            city={Mardan},
            postcode={23200}, 
            state={Khyber Pakhtunkhwa},
            country={Pakistan}}

            \author{Ibrar Ali Shah} 
\affiliation{organization={Department of Computer Software Engineering, University of Engineering and Technology},
            city={Mardan},
            postcode={23200}, 
            state={Khyber Pakhtunkhwa},
            country={Pakistan}}
            
\author{Farzana Zahid} 
\affiliation{organization={Department of Computer Science, University of Waikato},
            city={Hamilton },
            postcode={3240}, 
            country={New Zealand}}
\author{Daniyal Munir} 
\affiliation{organization={WICON, RPTU University, Kaiserslautern-Landau},
            city={Kaiserslautern},
            postcode={67663}, 
            country={Germany}}             
\author{Hans D. Schotten} 
\affiliation{organization={WICON, RPTU University, Kaiserslautern-Landau},
organization={ and Department of Intelligent Networks, German Research Center for Artificial Intelligence (DFKI)}, 
            city={Kaiserslautern},
            postcode={67663}, 
            country={Germany}}

\begin{abstract}

Security in cloud computing has become a major concern, as it is a core off-premises platform that provides services to numerous organisations while managing their sensitive information. There are several factors that are involved in complicating the security of the cloud environment, such as the layered architecture of the cloud, the cloud dynamic environment, and dealing with unseen or zero-day attacks. Moreover, mainly \ac{IDS} often deal with the security in a specific layer and are exclusively relying on machine learning models, which often show strong results in experimental setups but fail to maintain the same level of performance when deployed in real cloud environments. In our proposed work, we implement a confidence-aware multilevel intrusion detection system using a reinforcement learning designed for a cloud environment. This provides promising results for securing three distinct cloud layers: network, host, and hypervisor. Machine learning models are employed at each layer to identify known attack patterns, while prediction confidence is explicitly used to distinguish reliable decisions from uncertain cases. In the applied multi-gate flow, low-confidence events first pass through a learned-threshold confidence gate (Gate-1), then a Chroma memory-matching gate (Gate-2), and unresolved cases are escalated to a \ac{LLM} for semantic analysis and explanation. Final attack promotion at Gate-3 is performed using either direct (calibrated) \ac{LLM} confidence thresholds or weighted-fusion fallback (for borderline cases), while non-promoted uncertain events are retained in a low-confidence review bucket to avoid forced classifications. The generated explanations and confirmed attack knowledge are persistently stored in the attack vector database (ChromaDB) to support future analysis and incremental retraining. The approach is initially implemented across the three \ac{IDS} layers using a static confidence threshold, which sets the baseline later used to evaluate against the proposed pipeline. The results demonstrate that the proposed pipeline overcomes key limitations of static confidence implementations by autonomously learning layer-specific thresholds and reducing unnecessary \ac{LLM} escalations by 58.78\%, thereby lowering operational inference cost while maintaining strong end-to-end performance (88.68\% accuracy, 85.29\% precision, 84.72\% recall, and 85.00\% F1). The network and hypervisor layers achieved accuracies of 98.02\% and 97.08\%, respectively, resulting in a balanced, resource-efficient system with reliable detection.
\end{abstract}

\end{frontmatter}


%
\section{Introduction}
\label{sec:Intro}
Cloud computing has progressively replaced traditional on-premises infrastructures by providing scalable, on-demand computing resources, rapid deployment capabilities, and cost-effective infrastructure management over the internet \citep{baladari2021role,othman2024cloudIDS}. 
This widespread adoption enables organisations to depend on cloud environments for core business operations and the storage of sensitive data, thereby making these platforms essential for ensuring operational continuity and maintaining service availability.  \citep{arogundade2023virtualization,akhtar2021comprehensive,alzahrani2024cloudIDS}.
Reports from Gartner indicate that cloud computing is expected to become a business necessity by 2028, and market forecasts further suggest that the cloud market will double in size within the next eight years \citep{Donat2025}. 

Cloud computing offers significant benefits to organisations and customers; however, its shared, multi-tenant, and multi-layered architecture expands the attack surfaces, making it vulnerable to various cyber threats \citep{hashim2024securing}. Each architectural layer presents distinct security concerns, such as network-level attacks involving abnormal traffic patterns, host-level attacks arising from system misuse and suspicious activities, and sophisticated attacks, like \ac{VM} escape, that target the virtualisation layer by exploiting hypervisor vulnerabilities, which can affect multiple \ac{VM}s simultaneously  \citep{ahmed2023cloudIDS, Ahmadi2024, li2023hypervisorSecurity}. To address these security concerns, cloud-based \ac{IDS} are widely used to monitor network traffic (\ac{NIDS}), system activities (\ac{HIDS}), and virtualisation environments for each architectural layer \citep{ahmed2023cloudIDS}. 

Generally, \ac{IDS} operating at a single layer often fails to detect attacks that span across other layers, highlighting the need for proactive multi-layer intrusion detection mechanisms to improve cloud security \citep{li2023hypervisorSecurity}. To improve the performance of cloud-based \ac{NIDS} and \ac{HIDS},  \ac{ML} techniques have been widely reported \citep{liu2019machine,ahmad2021intrusion, alzahrani2022cloud, zhang2023deepcloud}.
However, these \ac{IDS} still face challenges as evolving zero-day attacks and unrealistic evaluation strategies lead to overly optimistic results and poor generalisation  \citep{ferrag2023cloudids,aliferis2024overfitting}.

Recently, \ac{LLM}s have emerged as powerful tools for attack analysis and detection in cloud environments, enabling intelligent automation across threat detection, vulnerability assessment, and incident response \citep{basiouni2025context,xu2024llm4security}. The ability of \ac{LLM}s to understand the context and interpret complex patterns makes them particularly well-suited for cloud-based \ac{IDS}, where they can perform tasks such as anomaly detection, network intrusion identification, and real-time threat classification more effectively than traditional rule-based or signature-based approaches \citep{AlSenani2025survey}. In practice, \ac{IDS} are now being integrated into security operations centre (SOC) workflows through platforms such as Microsoft Security Copilot, where they assist with phishing analysis, threat intelligence summarization, and real-time incident response support \citep{AgenticAI2026}. More recently, agentic LLM systems have further extended these capabilities by autonomously coordinating multi-step security operations across the entire incident lifecycle, from risk assessment and detection to response and post-incident recovery \citep{SpringerLLMCyber2025}. 

Our survey (Section~\ref{sec:Lit}) shows that despite growing interest in \ac{ML} and \ac{LLM}-assisted intrusion detection for cloud environments, most existing studies focus on a single detection layer and do not consider how network, host, and hypervisor layers collectively secure and operate within a unified detection system \citep{xu2024llm4security, Watanabe2024}. In addition, current \ac{IDS} pipelines rely on static confidence thresholds that do not adapt to the different behavior of these layers. This limitation can result into two key issues. First, it may cause unnecessary escalation, where events that are uncertain at a specific layer-level classifier (i.e., with confidence below the learned threshold) are forwarded to the next stage instead of being resolved at the current layer. This often results in additional analysis by \ac{LLM}s, increasing computational cost. Second, it can lead to insufficient escalation, where events that should be further analyzed are not forwarded, potentially resulting in missed attacks \citep{basiouni2025context}. 

Although \ac{LLM}s provide strong contextual reasoning for analysing uncertain and zero-day attacks, their uncontrolled use can reduce detection efficiency and increase operational overhead \citep{AlSenani2025survey}. Furthermore, existing systems lack mechanisms for deciding whether alerts should be handled internally or escalated to \ac{LLM} analysis based on model confidence, and they do not integrate historical attack memory to resolve borderline cases \citep{SpringerLLMCyber2025}. To address the limitations of static decision systems, recent research has explored \ac{RL} as an adaptive decision-making approach. In the \ac{RL} approach, the system learns from its experiences and continuously adjusts its decisions in response to dynamic environments \citep{Louati2024RL}. In particular, deep Q-learning (a type of \ac{RL}) provides continuous learning through trial-and-error interactions, enabling improved intrusion detection performance over time \citep{Alavizadeh2022DQL}. Studies have also shown that \ac{RL}-based models with online uncertainty estimation outperform static threshold models in heterogeneous environments, supporting the use of adaptive threshold learning for multi-layer \ac{IDS}s \citep{HATSRL2026}. \textit{Therefore, there remains a clear gap in developing a comprehensive cloud \ac{IDS} framework that integrates multi-layer detection, adaptive threshold-based routing, memory-assisted decision support, and \ac{LLM}-based contextual analysis within a single system.} 
\begin{figure*}[!t]
    \centering
    \includegraphics[width=\textwidth]{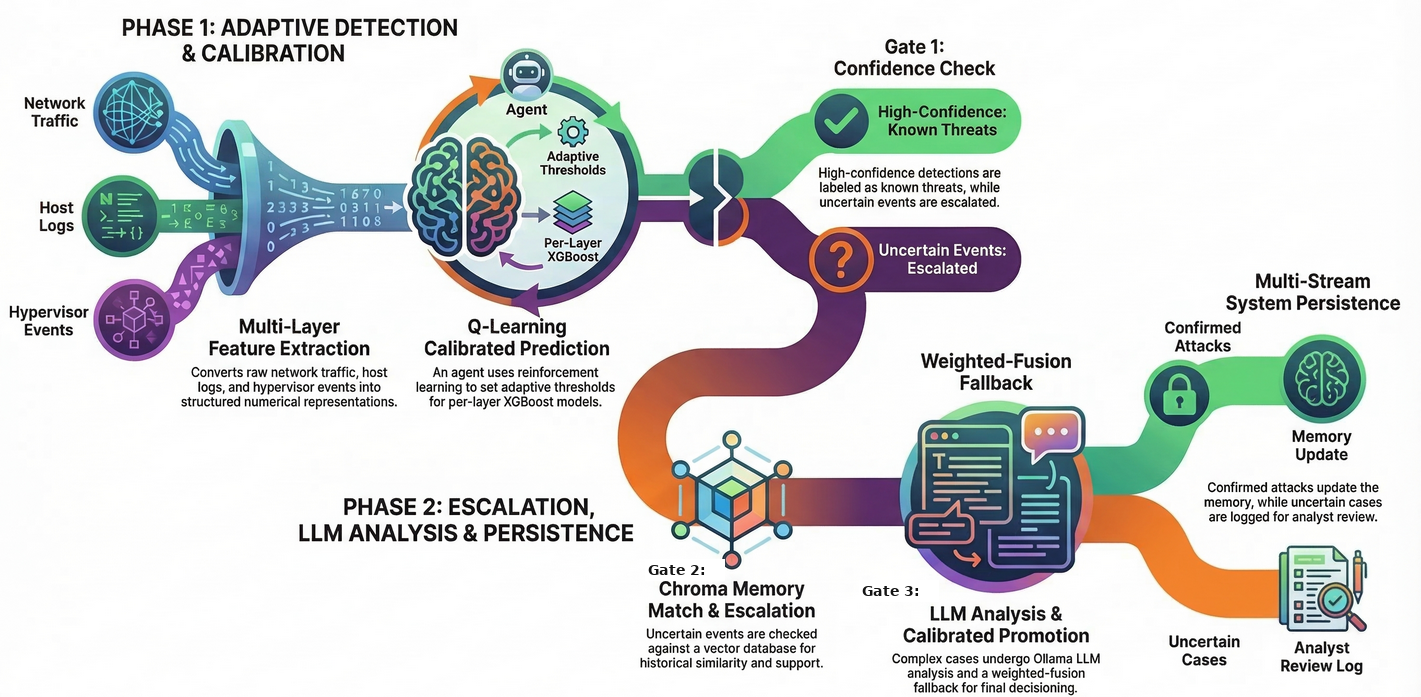}
    \caption{High-level Architecture of Proposed Multi-Layer Cloud \ac{IDS}  }
    \label{fig:cloudarch}
\end{figure*}

To address the above-mentioned issues, we proposed a multi-layer cloud \ac{IDS} framework (Fig.\ref{fig:cloudarch}). The architecture of the proposed \ac{IDS} comprises two phases: \textit{1) Adaptive detection and calibration, and 2) Escalation \ac{LLM} analysis and persistence}. The first phase performs adaptive detection using \ac{XGBoost} model with Q-learning-based threshold calibration, while the second phase handles low-confidence events through memory matching, \ac{LLM}-based analysis, and persistent logging with feedback-driven retraining. The framework is evaluated under two pipeline implementations: a static threshold-based \ac{IDS} pipeline, which serves as a baseline, and the proposed adaptive \ac{IDS} pipeline, both described in detail in Section~\ref{framework}.
\textit{The main contributions of this paper} are summarised as follows:

\begin{enumerate}
    \item A unified multi-layer cloud-based intrusion detection framework that integrates network, host, and hypervisor-level monitoring within a single detection architecture.
    
    \item An \ac{RL}-based adaptive threshold mechanism to automatically learn layer-specific confidence thresholds for intrusion detection decisions.
    
    \item An uncertainty-aware intrusion detection strategy that integrates \ac{ML} detection with \ac{LLM}-assisted analysis for low-confidence events.
    
    \item A memory-assisted attack knowledge mechanism for event logging, analysis, and future detection improvement.
    
    \item  A comparative evaluation with existing intrusion detection approaches and between static and adaptive decision mechanisms to demonstrate the effectiveness of the proposed framework.
\end{enumerate}
The remainder of this paper is organized as follows. 
Section~\ref{sec:Lit} presents the literature review on cloud \ac{IDS} and related \ac{ML}-based approaches. 
Section~\ref{framework} describes the proposed multi-layer intrusion detection framework and adaptive decision mechanism. 
Section~\ref{imp} presents the implementation details and discussion of the proposed system. 
Section~\ref{res} provides the results evaluation and performance analysis. 
Finally, Section~\ref{con} concludes the paper and outlines future research directions.
\section{Literature Review}
\label{sec:Lit}

In the literature, several studies have explored \ac{IDS} using \ac{DL} and classical \ac{ML} approaches. 
The study \citep{Long2024} evaluated a transformer-based \ac{NIDS} using the Canadian institute for cybersecurity intrusion detection system 2018 (CICIDS2018) dataset for traffic-based attacks such as \ac{DoS}, distributed \ac{DoS} (D\ac{DoS}), botnet, and infiltration.  Their model achieved an accuracy of up to 99\% for DoS and botnet attacks, but showed poor performance for infiltration and brute force attack patterns in cloud network environments.
Similarly, in \citep{Joraviya2024DLHIDS}, researchers proposed a \ac{DL}-based \ac{HIDS} for containerised cloud environments using system call monitoring at the kernel level, and the Leipzig intrusion detection dataset (LID-DS-2019).

It has also been observed that \ac{XGBoost} is widely used for the development of \ac{IDS} due to its high detection performance and scalability for cloud environments. In \citep{Zatika2024XGBoostIDS}, a real-time \ac{XGBoost}-based \ac{IDS} with hyperparameter optimisation has been proposed that improved detection performance and demonstrated low computational overhead. Similarly, \citep{XGBoostIDS2017} proposed an \ac{XGBoost}-based \ac{IDS} classification model that achieved 98.7\% accuracy and outperformed several traditional \ac{ML} models. These studies highlight \ac{XGBoost} as an efficient and scalable approach for cloud-based \ac{IDS}.

Moreover, hybrid learning models have been explored for cloud-based \ac{IDS} using a convolutional neural network-recurrent neural network (CNN-RNN) model evaluated on the network security laboratory-knowledge discovery in databases (NSL-KDD) dataset \cite{9461285}, demonstrating effective spatial and temporal traffic pattern learning. The study \citep{Geetha2024} proposed the optimized bidirectional convolutional long short-term memory (OBCLSTM (Bi-CNN-LSTM)) model evaluated on multiple datasets, including NSL-KDD99, Ton Duc Thang University intrusion detection system (TUIDS), University of New South Wales-network benchmark 2015 (UNSW-NB15), and botnet of things-internet of things (BoT-IoT), achieving a 99.1\% detection rate and highlighting model generalizability for cloud environments. Additionally, a deep autoencoder combined with CNN was used for low-rate DDoS detection \cite{10179834}, achieving 95.32\% accuracy by leveraging unsupervised feature learning for detecting stealthy cloud-based attacks. Collectively, these studies show that hybrid learning models perform well for frequent and well-represented attack types, but their performance decreases for rare or new attacks, highlighting the need for adaptive and multi-layer intrusion detection frameworks.

Recent studies have started integrating \ac{LLM}s into \ac{IDS} to handle uncertain or ambiguous cases that traditional models cannot effectively classify. For example, a two-tier \ac{LLM}-based \ac{IDS} is proposed by \citep{Kalafatidis2025} for software-defined networking (SDN) and Kubernetes environments that combine statistical anomaly detection methods such as exponential moving average (EMA) and auto-regressive integrated moving average (ARIMA) with on-demand packet capture. The system was evaluated using real attack traffic, including user datagram protocol (UDP) floods, port scans, synchronize (SYN) floods, secure shell (SSH) brute-force, slowloris, and HTTP/2 attacks. Results showed faster detection in SDN (1-1.5s) compared to Kubernetes (6–30s), but the approach introduced response delays due to application programming interface (API) communication and \ac{LLM} inference time. Whereas, in \citep{adjewa2025llm}, an \ac{IDS} based on transformer language models was introduced for detecting evolving and unknown network attacks through continuous learning.
Also, researchers in \citep{Tavallaee2024HybridNIDSHIDS} developed a hybrid intrusion detection framework combining network and host-based features using a two-stage collaborative classifier to improve detection accuracy and reduce computational complexity.

Similarly, a federated \ac{IDS} framework was developed in \citep{FernandezSaura2026FGCS} where low-confidence alerts from the base detector are forwarded to an \ac{LLM} for secondary analysis using external threat intelligence and a federated \ac{RAG}-based incident knowledge base. The system also feeds \ac{LLM}-verified alerts back into the training process to improve detection. Results showed improved recall for underrepresented attacks and reduced false positives, but the approach still relies on fixed confidence thresholds and lacks a calibrated decision mechanism for borderline \ac{LLM} cases. Furthermore, \citep{Blefari2507} proposed CyberRAG, an agentic \ac{RAG}-based system combining a central \ac{LLM} with three fine-tuned transformer classifiers for structured query language (SQL) injection, cross-site scripting (XSS), and server-side template injection (SSTI) attacks, achieving 94.92\% accuracy with \ac{RAG} compared to 84.75\% without \ac{RAG}. The basic limitation of the study is dependency on predefined classifiers and static knowledge bases, restricting generalisation to novel attack types.

Intrusion detection at the hypervisor layer has also gained increasing attention.
Several studies have explored \ac{ML}, federated learning, bio-inspired optimisation, and \ac{RL} approaches for detecting hypervisor-level attacks such as hyperjacking, side-channel attacks, and \ac{VM} escape attacks \citep{Jaber2020, Alazab2025, 10503284, Qaffas2024}. These studies demonstrate that hypervisor-level monitoring can significantly enhance cloud security by detecting virtualisation-specific threats. However, most existing approaches focus on isolated hypervisor detection mechanisms and do not integrate hypervisor monitoring within a unified multi-layer \ac{IDS} framework, highlighting the need for integrated multi-layer cloud intrusion detection architectures.

Since attacks evolve with time, their detection also evolves. However, new detection tools need to be adaptive to such evolving attacks. In this regard, \ac{RL} has recently presented itself as a natural candidate for adaptive \ac{IDS} to overcome the static nature of traditional models. The study in \citep{Hossain2025DQIDS} developed a deep q-network (DQN)-based security mechanism for achieving high accuracy and low computational overhead; while \citep{MogollonGutierrez2026RLIDS} used \ac{RL} to optimise classifier ensemble decisions, improving detection performance. In general, these studies show that \ac{RL} enables adaptive decision-making and dynamic optimisation in \ac{IDS}.

The reviewed literature highlighted several limitations, including scalability constraints, high \ac{LLM} escalations, reliance on curated knowledge bases, insufficient hypervisor-level threat coverage, static threshold configurations, and evaluation restricted to single datasets, collectively exposing a clear research gap. \textit{Overall}, our work addresses these limitations through a comprehensive \ac{IDS} pipeline for cloud environment. The pipeline is built using an \ac{XGBoost} classifier with adaptive \ac{RL}-based thresholding and \ac{LLM}-assisted decision analysis, combined with precision-constrained calibration of detection thresholds and weighted fusion for handling borderline cases. The proposed framework adopts a cloud-oriented multi-layer \ac{IDS} architecture with chroma-supported memory to accumulate and reuse attack knowledge.
\section{Proposed Framework}
\label{framework}
This section presents the proposed multi-layer cloud intrusion detection framework (Fig. \ref{fig:cloudarch}) for security monitoring across network, host, and hypervisor layers in cloud environments.
The proposed adaptive \ac{IDS} pipeline is divided into two primary phases. The following subsections discuss the functionality of each phase separately. 
\subsection{Phase 1 (Adaptive Detection and Calibration)} 
\begin{figure*}[!t]
    \centering
    \includegraphics[width=\textwidth]{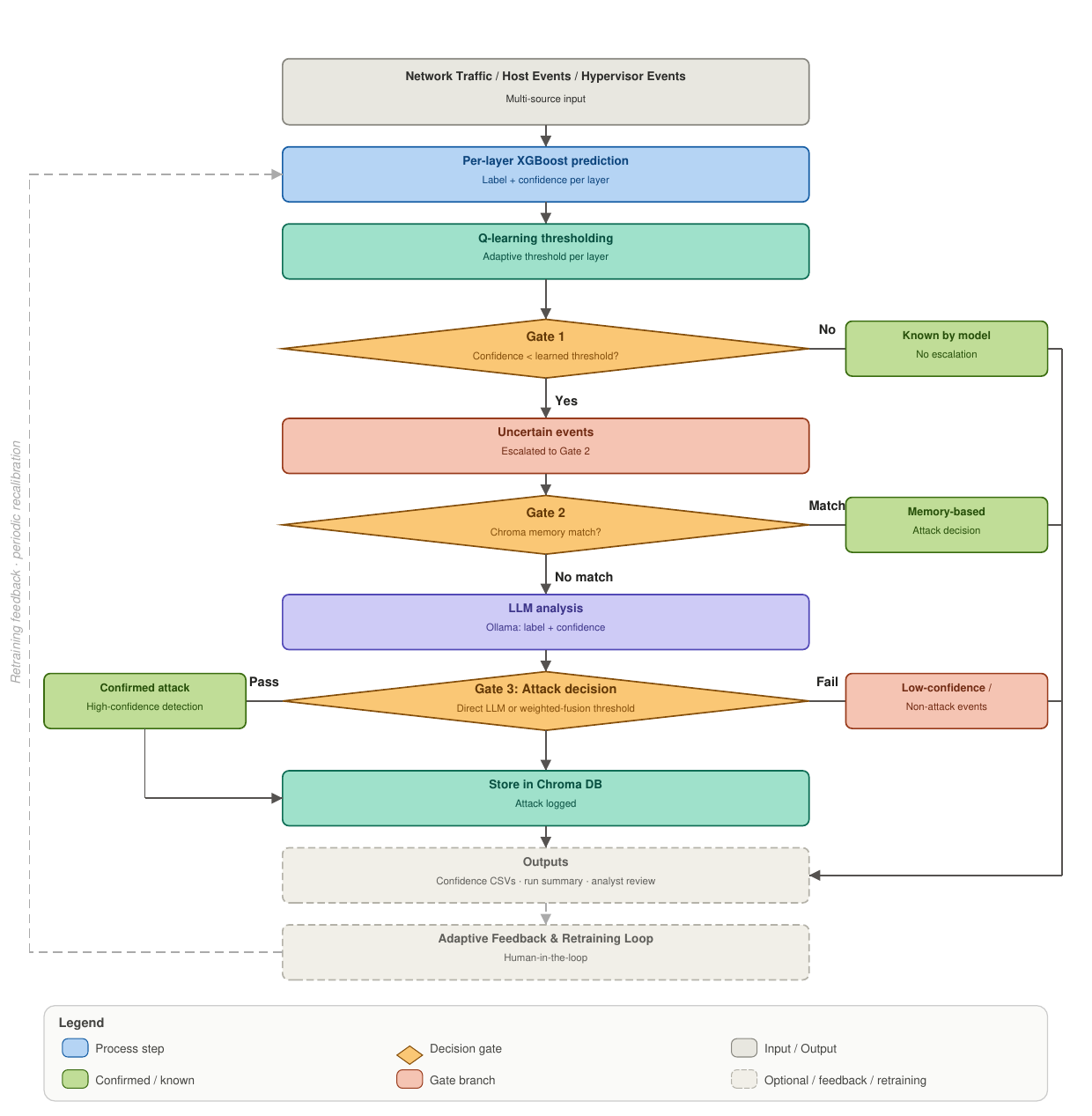}
    \caption{Proposed Q-learning-calibrated  IDS pipeline with three 
    adaptive decision gates, Chroma vector memory, LLM analysis (Ollama), 
    and a periodic retraining feedback loop.}
    \label{fig:ids_proposed}
\end{figure*}

In Phase 1, we propose a Q-learning-based cloud intrusion detection pipeline that serves as an agent for autonomous confidence-threshold selection. As shown in Fig. \ref{fig:ids_proposed}, the framework integrates XGBoost with an \ac{LLM} and memory pipeline to support detection of both known attacks (previously observed threats with identifiable signatures or patterns) and unknown attacks (novel threats that do not match any previously seen or learned pattern) within the three-layer \ac{IDS}. 
\ac{XGBoost} is selected as the base classifier due to its high classification accuracy, computational efficiency, and robustness on tabular network data \citep{gouveia2020network, shwartz2022tabular}.

Phase 1 consists of two main processes: \textit{multi-layer feature extraction} and \textit{calibrated prediction}.
In the first process, raw data is collected from each layer, including network traffic, host logs, and hypervisor events, which is transformed into structured numerical representations using layer-specific feature extraction techniques. For the network layer, flow-based statistical features from the CICIDS2018 dataset are utilized. For the host layer, textual logs are converted into numerical vectors using term frequency-inverse document frequency (TF-IDF). For the hypervisor layer, event-based logs are parsed and encoded into structured representations.

In the second process, \textit{calibrated prediction}, the next step is per-layer \ac{XGBoost} prediction (labels with confidence scores where the confidence is defined as the class probability). This is followed by Q-learning-based adaptive thresholding, which determines an optimal confidence threshold for each layer by iterative learning, based on prediction outcomes. The Q-learning threshold model was trained on twenty episodes of events to achieve stable threshold decisions. The number of episodes is decided based on the nature of the problem, as the Q-learning model only learns to adjust confidence thresholds, not complex decision strategies, so that it can reach stable threshold decisions \citep{sutton2018reinforcement}. For each layer, the Q-learning model selects a threshold from a predefined threshold set, uses it to route each event as KNOWN (confidence is equal to or greater than the learned threshold) or UNCERTAIN (confidence is less than the learned threshold), receives reward based on correctness/risk/escalation cost and uncertain-band control, and updates Q-values with the Bellman rule as shown in eq.~(\ref{eq:qlearning bellman}), originally introduced in the Q-learning framework \citep{watkins1992qlearning}. 
\begin{equation}
Q_{t+1}(s_t,a_t)=Q_t(s_t,a_t)+\alpha\left[r_t+\gamma \max_{a'}Q_t(s_{t+1},a')-Q_t(s_t,a_t)\right]
\label{eq:qlearning bellman}
\end{equation}
where:
\begin{itemize}
\item $s_t$ is the current state (discretized confidence statistics),
\item $a_t$ is the selected threshold action,
\item $r_t$ is the reward from the routing decision,
\item $\alpha$ is the learning rate, and
\item $\gamma$ is the discount factor.
\end{itemize}

After the episodes finish, it runs a greedy rollout to pick the most frequently selected threshold as the learned threshold for Gate 1, as shown in Fig \ref{fig:ids_proposed}. It continuously monitors confidence scores, uncertainty indicators, and alert patterns produced by the \ac{NIDS}, \ac{HIDS}, and hypervisor layers. Based on the observed system behavior, the Q-learning mechanism dynamically adjusts confidence thresholds to balance detection reliability and alert overhead. In other words, it enables each \ac{IDS} layer to decide for itself whether it is confident enough to classify the event and when to escalate to \ac{LLM} for assistance. The output of this mechanism reduces unnecessary escalation to \ac{LLM} while staying reliable under different conditions. 

The model’s state representation captures short-term confidence statistics, including the mean confidence, confidence variance, and the ratio of uncertain events over a sliding window, while actions correspond to selecting confidence thresholds from a predefined range. A reward mechanism is designed to encourage stable detection outcomes, control uncertainty, and use computational resources efficiently. In particular, correct high-confidence predictions receive positive rewards, whereas misclassifications and unnecessary \ac{LLM} escalations are penalized. Through iterative interaction with the detection pipeline, the Q-learning agent (RL-based threshold calibration mechanism) gradually learns an effective threshold adjustment policy that optimizes long-term routing performance without modifying the underlying detection models. This process minimizes uncertainty cost, defined as the computational overhead and response delay caused by escalating ambiguous events for advanced \ac{LLM}-assisted analysis, as shown in Fig.~\ref{fig:ids_proposed}. During this phase, Gate-1 determines whether an event should be directly accepted or escalated. If the confidence exceeds the learned threshold, the event is classified as a high-confidence prediction (no escalation); otherwise, the low-confidence event proceeds to Gate-2 in Phase 2.

\subsection{Phase 2 (Escalation, LLM Analysis, and Persistence)}
Phase 2 includes three processes:  \textit{escalation}, \textit{\ac{LLM} analysis}, and \textit{persistence}.  
The escalation process starts with a memory-based matching gate, called Gate 2, where each escalated uncertain event (during Phase 1) is compared against the Chroma vector memory. If a similarity is found, defined as the semantic similarity between the embedding of the current event and the embeddings of previously confirmed attack events stored in memory (referred to as known attack patterns), the event is classified as an attack through a memory-based decision; otherwise, it is forwarded to Gate-3 for \ac{LLM}-based final validation. 

The memory match is based on three factors: distance, support, and meta-confidence. The distance measures semantic similarity, support shows how many consistent neighboring attack records back the match, and meta-confidence summarizes the reliability of that memory evidence, where the current event embedding is compared against previously confirmed attack embeddings \citep{liao2002use, lewis2020retrieval}.
Initially (first execution cycle of the pipeline), the Chroma memory is empty, as no previously confirmed attack embeddings are available. Therefore, most events bypass Gate-2 and proceed to the \ac{LLM} analysis stage. With the confirmed attacks stored over time, memory becomes progressively populated, enabling Gate-2 to directly classify events as attacks through memory-based decisions for high-confidence matches. The events with low similarity or no reliable match are forwarded to \ac{LLM} analysis for contextual re-evaluation.

The next process in Phase 2 is \ac{LLM} analysis (Ollama), followed by Gate 3. Gate 3 is a final decision gate that uses \ac{LLM} confidence to make the final classification decision. At Gate 3, classification follows two calibrated rules. First, a direct rule is used: an event is accepted as ``ATTACK'' only when the \ac{LLM} predicts ATTACK and its confidence exceeds a layer-specific threshold learned via precision-constrained calibration (not a fixed heuristic as used in recent \ac{LLM}-based \ac{IDS} approaches discussed in Section~\ref{sec:Lit}). The precision constraint is defined as: 
\begin{equation}
\label{Prec}
\mathrm{Precision}(t) \geq P_{\min},
\end{equation}
where $P_{\min} = 0.80$ is set to ensure an operationally manageable false-positive rate (i.e., $\geq 80\%$ precision), acknowledging that excessive false alarms undermine practical deployment \citep{axelsson2000baseratefallacy, soc_alert_fatigue_acm2025}. The decision threshold is then chosen to maximize recall under this constraint. Among all thresholds satisfying this constraint, the selected threshold is the one that preserves the strongest detection coverage, which can be written as,
\begin{equation}
\label{Rec}
\begin{aligned}
t^*= & \quad \arg\max_t \mathrm{Recall}(t)\\
 \text{subject to:}& \quad
\mathrm{Precision}(t)\geq P_{\min},
\end{aligned}
\end{equation}

The decision threshold is then chosen to maximize recall under this constraint. Among all thresholds satisfying this constraint, the selected threshold is the one that preserves the strongest detection coverage, which can be written as,
\begin{equation}
\label{Rec2}
\begin{aligned}
t^*= & \quad \arg\max_t \mathrm{Recall}(t)\\
 \text{subject to:}& \quad
\mathrm{Precision}(t)\geq P_{\min},
\end{aligned}
\end{equation}
During escalation, the \ac{LLM} assigns a semantic decision label together with a confidence score. If the confidence exceeds the learned layer-specific threshold, the corresponding decision is accepted; otherwise, the event is marked as UNSURE (Which are the low confidence events) and routed for further analysis. The direct \ac{LLM} decision rule can be mathematically expressed as,
\begin{equation}
\label{eq:llm_decision}
\hat{y}_{\mathrm{LLM}} =
\begin{cases}
1, & \text{if } d_{\mathrm{LLM}}=\mathrm{ATTACK}
\land c_{\mathrm{LLM}} \ge \tau_{\mathrm{LLM},\ell}, \\

0, & \text{if } d_{\mathrm{LLM}}=\mathrm{BENIGN}
\land c_{\mathrm{LLM}} \ge \tau_{\mathrm{LLM},\ell}, \\

\mathrm{UNSURE}, & \text{if } c_{\mathrm{LLM}} <
\tau_{\mathrm{LLM},\ell}.
\end{cases}
\end{equation}
where \(d_{\mathrm{LLM}}\) denotes the semantic decision generated by the \ac{LLM}, \(c_{\mathrm{LLM}}\) represents the \ac{LLM} confidence score, and \(\tau_{\mathrm{LLM},\ell}\) is the learned confidence threshold for layer \(\ell\).

Second, if the direct \ac{LLM} confidence falls below this threshold, these borderline cases (events predicted with confidence slightly below the threshold) are not discarded; instead, they are processed using a weighted-fusion fallback mechanism for event classification. A weighted-fusion fallback is applied by linearly combining base-model confidence and \ac{LLM} confidence using layer-specific calibrated weights and a calibrated fusion threshold \citep{guo2017calibration, kittler1998combining}. If the resulting fused score meets the same \ac{LLM} confidence threshold criterion used in the direct decision rule, the event is classified as a confirmed attack; otherwise, it is routed to the low-confidence(UNSURE)/non-attack JSONL review bucket, as shown in Fig~\ref{fig:ids_proposed}. A weighted-fusion fallback is applied when the \ac{LLM} predicts ATTACK but its confidence is below the direct per-layer \ac{LLM} threshold. We apply fusion only to uncertain cases, where the classifier is already weak. Therefore, the \ac{LLM} receives a higher weight (0.8) to lead the decision, while the classifier retains a smaller weight (0.2) as a safety check, with layer-specific fusion thresholds aligned to each layer’s \ac{LLM} threshold. This provides a calibrated and transparent justification basis for borderline decisions. A weighted-fusion fallback can be given as,
\begin{equation}
S_{\mathrm{fusion}} = w_m\,C_{\mathrm{model}} + w_l\,C_{\mathrm{LLM}}, 
\end{equation}
where \(C_{\mathrm{model}}\) is the base classifier confidence, \(C_{\mathrm{LLM}}\) is the LLM confidence, and $w_m$ and $w_l$ are weights assigned to classifier and \ac{LLM}, respectively, which satisfy \(w_m + w_l = 1\). The fallback decision rule can be mathematically written as,
\begin{equation}
\label{Fallback}
\begin{aligned}
\hat{y} =
\begin{cases}
1 
\quad\quad (\mathrm{LLM\ label}=\mathrm{ATTACK}) \land \left(S_{\mathrm{fusion}} \ge \tau_{\mathrm{fusion},\ell}\right), \\
0 
\quad\quad \text{otherwise,}
\end{cases}
\end{aligned}
\end{equation}

where \(\tau_{\mathrm{fusion},\ell}\) denotes the layer-specific fusion threshold, which is set equal to the corresponding layer's \ac{LLM} attack threshold.

In addition, specific fusion weights used in this study ($w_m = 0.20$, $w_l = 0.80$) were selected empirically based on the validation behavior of the uncertain-event subset. This choice is consistent with prior work showing that weighted score fusion is a valid approach for combining heterogeneous classifier outputs and confidence scores \citep{xu2015adaptive, nguyen2020confidence}.

The post-gate stages perform attack persistence by storing confirmed attack events (i.e., events classified as attacks at Gate-3) in the Chroma attack database (ChromaDB), uncertainty auditing by logging Gate-1 uncertain cases, and output finalization by generating confidence Comma-Separated Values (CSVs), run summaries, and analyst review records \citep{ortega2022taxonomy, abdar2021review}. The review bucket is retained as structured JSONL, so that unresolved events remain traceable and auditable rather than being discarded. These persistent outputs close the loop by updating the knowledge base for the future retraining stage, where accumulated validated events are used for periodic recalibration of thresholds/weights and for retraining an independent \ac{XGBoost}-based classifier to maintain detection performance over time.
\begin{figure}[!t]
    \centering
    \includegraphics[width=\columnwidth]{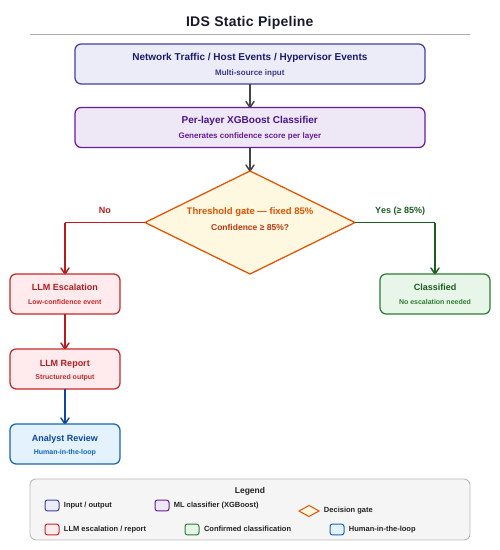}
    \caption{Static thresholding baseline IDS pipeline using a fixed 85\% 
    XGBoost confidence threshold for intrusion detection.}
    \label{fig:ids_baseline}
\end{figure}

To provide a comparative baseline for evaluating the proposed adaptive framework, The study considers two pipeline implementations: the proposed adaptive \ac{IDS} pipeline based on \ac{RL}-driven threshold optimisation and uncertainty-aware decision routing, illustrated in Fig.~\ref{fig:ids_proposed}, and a static threshold-based three-layer \ac{IDS} pipeline implemented as a baseline reference framework, shown in Fig.~\ref{fig:ids_baseline}. The baseline pipeline represents a complete three-level security implementation comprising network, host, and hypervisor-based \ac{IDS}, each deployed on a separate layer. Both pipelines perform intrusion detection as a binary classification task, distinguishing between benign and attack activity across network, host, and hypervisor layers, enabling consistent and comparable evaluation under matched experimental conditions. The static threshold is fixed, as adopted from recent work \citep{FernandezSaura2026FGCS}, which applies a similar fixed-threshold approach. Fig \ref{fig:ids_baseline} illustrates the workflow of the static threshold-based three-layer \ac{IDS} pipeline, where the XGBoost classifier processes inputs from each layer. The classifier assigns a confidence score (the predicted probability that a given event is malicious or benign) to each flow. If the score is equal to or exceeds the static threshold, the event is classified, without escalation to the \ac{LLM}. On the contrary, if the confidence score falls below the static threshold, the event is escalated to the \ac{LLM} for further analysis. However, a fixed threshold fails to adapt to the varying traffic patterns across layers, leading to unnecessary \ac{LLM} escalations or missed detections.
Overall, the proposed framework integrates supervised detection with \ac{LLM}-based semantic reasoning, providing adaptive and explainable intrusion analysis across cloud network, host, and hypervisor layers, with \ac{RL}-driven threshold adaptation and persistent attack knowledge storage.
\section{Implementation and Discussion}
\label{imp}
The proposed multi-layer intrusion detection framework pipelines were implemented on a system with an Intel Core i5-8600K (6 cores), NVIDIA GeForce GTX 1080 GPU, 32 GB RAM, and 2.5 TB storage, using PyCharm 2025.3.1 as the development environment. The LLM component was implemented using Ollama. The \ac{XGBoost} classifier was implemented using the XGBoost 2.x series. For consistency, each layer is evaluated on the first 5,000 events from its held-out test split, which provides stable estimates while keeping runtime manageable for repeated experiments and \ac{LLM} escalation analysis.
\subsection{The Baseline (Static Confidence Threshold Based Three-Layer Cloud IDS Implementation)}
This subsection evaluates the implementation of the three-layer intrusion detection pipeline using a static confidence threshold. This implementation serves as a baseline for comparison with the proposed \ac{IDS} pipeline, where a fixed confidence threshold of 0.85 is applied across all layers (\ac{NIDS}, \ac{HIDS}, and hypervisor-\ac{IDS}), as described in Section~\ref{framework} and consistent with~\cite{FernandezSaura2026FGCS}. 
\subsubsection{Network-Based IDS (NIDS)} In this study, the publicly available CICIDS-2018 dataset \cite{sharafaldin2018toward} is used for the development of \ac{NIDS} as it provides realistic, labelled network traffic covering diverse attack categories including \ac{DoS}, DDoS, brute-force, and web attacks with rich flow-based features suitable for \ac{ML}-based network intrusion detection. To detect network traffic attacks, a binary classification setting is adopted, where all attack types in the dataset are grouped into a single \textit{Attack} class, while normal traffic represents the \textit{Benign} class (also described in Section~\ref{framework}). The \ac{XGBoost} model is trained using 40 features selected through feature-importance ranking and validation, providing a balance between detection accuracy and computational cost. Model performance is evaluated on unseen test data using standard metrics, including precision, recall, and F1-score.

\begin{figure}[t]
    \includegraphics[width=\columnwidth]{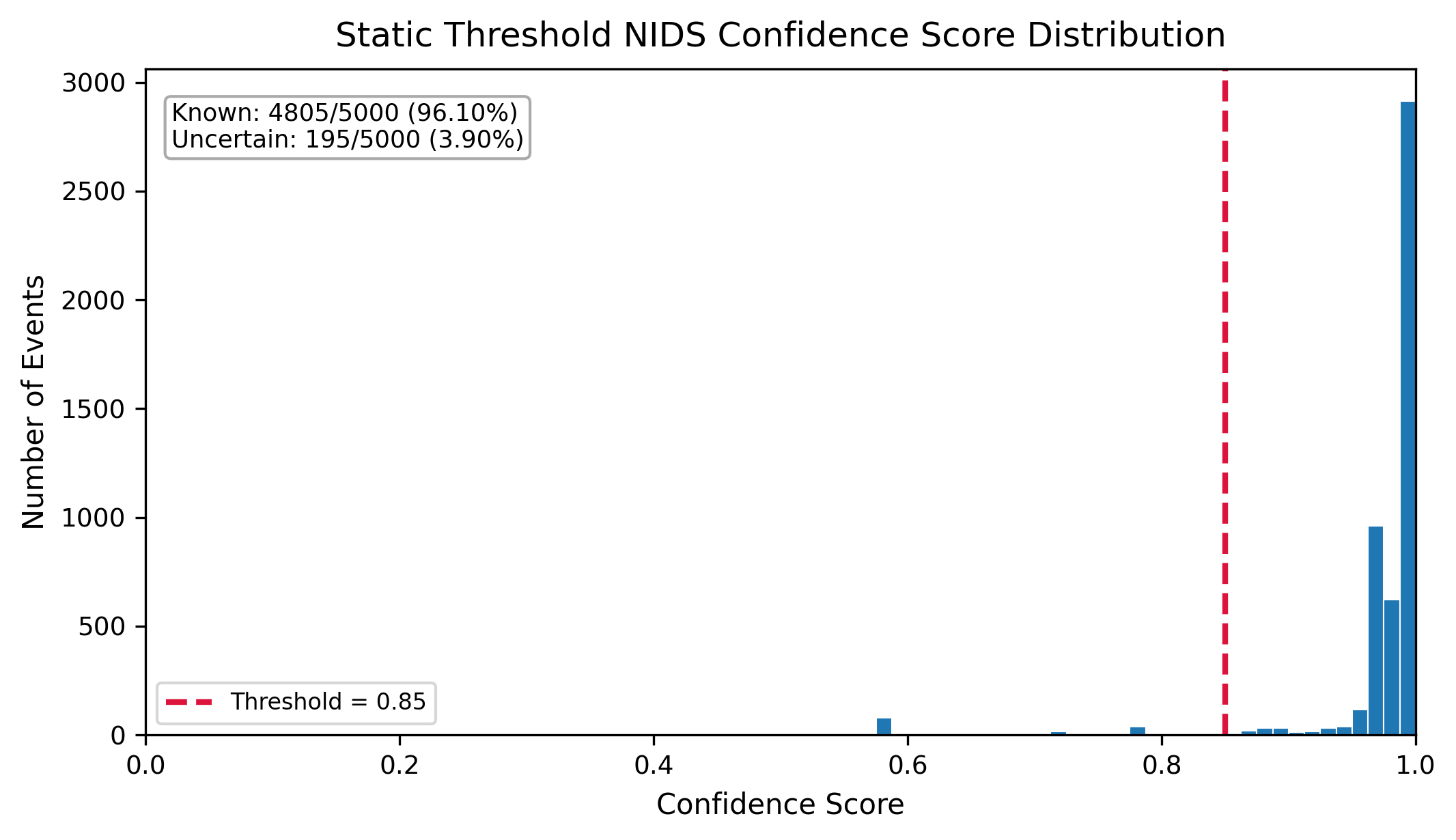}
    \caption{Distribution of prediction confidence scores produced by the NIDS component.}
    \label{fig:nidsconfidence1}
\end{figure}

For \ac{NIDS}, flows with confidence above the selected threshold (0.85) are treated as known events by \ac{XGBoost}, while those below the threshold are escalated to the \ac{LLM}. The confidence distribution and routing outcomes are shown in Fig.~\ref{fig:nidsconfidence1}. The red dashed vertical line denotes the confidence threshold. Bars to the right are high-confidence (known) predictions, and bars to the left are low-confidence (uncertain) predictions routed to the \ac{LLM}. In the evaluation subset of 5,000 flows, 4,805 flows (96.10\%) are classified as high-confidence and 195 flows (3.90\%) as low-confidence, based on model-generated confidence scores relative to the 0.85 threshold. These low-confidence and routed to the \ac{LLM}, based on threshold confidence scores. The example in Table~\ref{tab:nids_case} presents an uncertain \ac{NIDS} event extracted from the structured JSONL log file, which records all flows escalated to the \ac{LLM} during the 5,000-flow evaluation.

\begin{table}[ht]
\footnotesize
\setlength{\tabcolsep}{5pt}
\renewcommand{\arraystretch}{1.3}
\caption{Description of uncertain NIDS event promoted as ATTACK by LLM}
\label{tab:nids_case}
\begin{tabular}{|p{2.5cm}|p{5.5cm}|}
\hline
\textbf{Field} & \textbf{Description} \\
\hline
Event Type
& Suspected FTP Brute-Force Activity (NIDS) \\
\hline
ML Confidence Score
& 0.5807  \\
\hline
Routing Decision
& \textit{Uncertain} $\rightarrow$ LLM analysis \\
\hline
Observed Flow Entry
& \texttt{[NIDS FLOW] Dst Port=21, Protocol=6, Flow Duration=2, Tot Fwd Pkts=1, Tot Bwd Pkts=1, Flow Pkts/s=1000000, ...} \\
\hline
LLM-Based Interpretation
& The high packet rate and short flow duration suggest a potential brute-force attack on port 21, which is commonly used for FTP. \\
\hline
LLM Output
& label=\texttt{ATTACK}, confidence=\texttt{0.80}, attack\_type=\texttt{FTP brute force} \\
\hline
\end{tabular}
\end{table}
\subsubsection{Host-Based IDS (HIDS)}
To implement the host-based \ac{IDS} pipeline, experiments are conducted on the LID-DS 2019 dataset \cite{grimm2019lidds}, a benchmark for evaluating Linux \ac{HIDS} using system-call traces. This dataset is selected for its labeled host-level data that captures both normal and attack behaviors, aligning with the requirements of the host layer. Consistent with the unified setup across all layers, the classifier performs binary classification, predicting benign or attack classes. Events with confidence below 0.85 are labeled as "UNCERTAIN", whereas those with confidence greater than or equal to 0.85 are labeled as "KNOWN".

\begin{figure}[t]
    \includegraphics[width=\columnwidth]{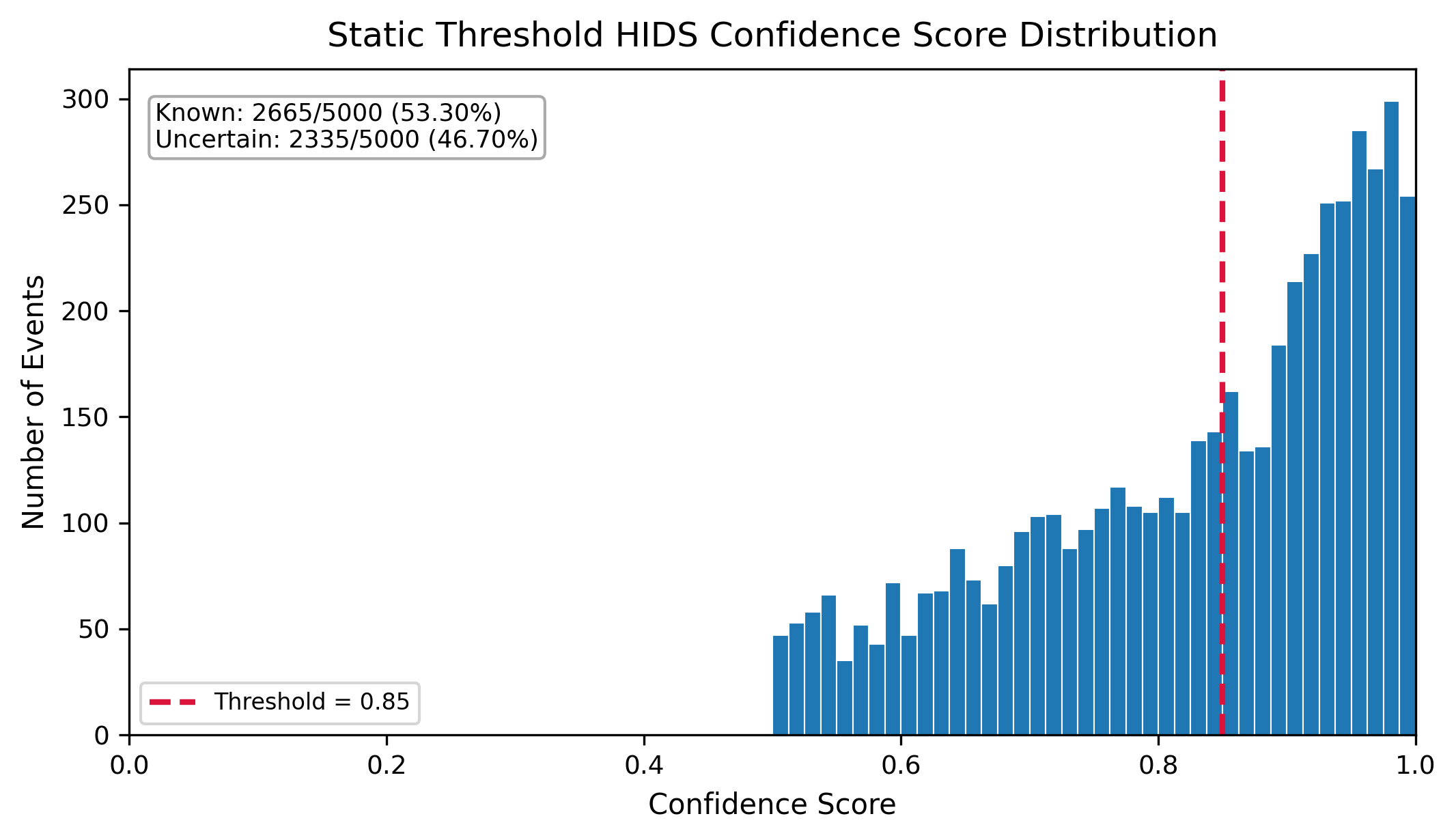}
    \caption{Distribution of prediction confidence scores produced by the HIDS}
    \label{fig:hidsconfidence1}
\end{figure}

For proper evaluation, a sample of 5,000 events is used to ensure stable estimates while maintaining computational efficiency for repeated evaluations and \ac{LLM} escalation analysis. Of these 5,000 events, 2,665 events (53.30\%) are classified as known, while 2,335 events (46.70\%) are marked as uncertain. The \ac{HIDS} prediction confidence score distribution, prediction outcome distribution, and details are shown in Fig~\ref{fig:hidsconfidence1}. Low-confidence events identified in Fig.~\ref{fig:hidsconfidence1} are marked as uncertain and forwarded to the \ac{LLM} layer for further analysis. The analyst-style explanation generated by the \ac{LLM} is then stored in ChromaDB to support improved future retrieval and reasoning. An example case scenario is presented in Table~\ref{uncertain HIDS}.
\begin{table}[ht]
\footnotesize
\setlength{\tabcolsep}{5pt}
\renewcommand{\arraystretch}{1.3}
\caption{Description of uncertain HIDS event via LLM}
\label{uncertain HIDS}
\begin{tabular}{|p{2.5cm}|p{5.5cm}|}
\hline
\textbf{Field} & \textbf{Description} \\
\hline
Event Type 
& Uncertain Brute-Force activity (CWE-307) \\
\hline
ML confidence score 
& 0.38 \\
\hline
Routing decision 
& \textit{Uncertain} $\rightarrow$ LLM analysis\\
\hline
Observed Log Entry 
& \texttt{apache2 ... > poll fds=12:41 timeout=5000} \\
\hline
LLM-Based Interpretation 
& The \texttt{apache2} process repeatedly enters a polling state with extended timeouts, which is atypical for normal web request handling. This behavior is commonly associated with automated clients probing authentication endpoints at a low rate to evade detection. Although a single log entry does not explicitly indicate credential failures, the execution pattern and timing characteristics suggest a potential brute-force probing attempt. Correlation with adjacent authentication logs is recommended for confirmation. \\
\hline
\end{tabular}
\end{table}
\subsubsection{Hypervisor based IDS}
In order to implement the complete \ac{IDS} pipeline, we have generated a synthetic hypervisor-based dataset for validation of our methodology. The dataset expresses a natural imbalance with a high ratio of noise, with 74\% benign and 26\% malicious events. Please note: despite an extensive search, we did not find any publicly available and labeled dataset for hypervisor-level intrusion detection. This is mainly due to the sensitive and privileged nature of hypervisor logs, which are rarely released by cloud providers. Therefore, we constructed a realistic synthetic hypervisor dataset to enable controlled and reproducible evaluation of our proposed approach. The summary of the generated hypervisor-based dataset is shown in Table~\ref{tab:hypervisor}.
\begin{table}[ht]
\centering
\caption{Synthetic Hypervisor Dataset Summary}
\label{tab:hypervisor}
\footnotesize
\setlength{\tabcolsep}{5pt}
\renewcommand{\arraystretch}{1.3}
\begin{tabular}{|p{2.5cm}|p{5.5cm}|}
\hline
\textbf{Property} & \textbf{Value} \\
\hline
Records 
& 25{,}000 \\
\hline
Hypervisor Type 
& VMware ESXi, KVM, Xen, Hyper-V (Type-1) \\
\hline
Target Layer 
& Hypervisor \\
\hline
Features / Columns 
& 24 total columns \\
\hline
Class Labels (for detection) 
& Binary ground truth: 2 classes (0=benign, 1=attack) \\
\hline
 Distribution of Data 
& normal: 12500; vm lateral-movement: 2541; vm escape: 2{,}501; snapshot abuse: 2500; hypervisor dos: 2488; hyper-jacking: 2470 \\
\hline
Pipeline Split (current) 
& Runtime random split in pipeline: 80/20 train/test \\
\hline
\end{tabular}
\end{table}

\begin{figure}[t]
    \centering
    \includegraphics[width=\columnwidth]{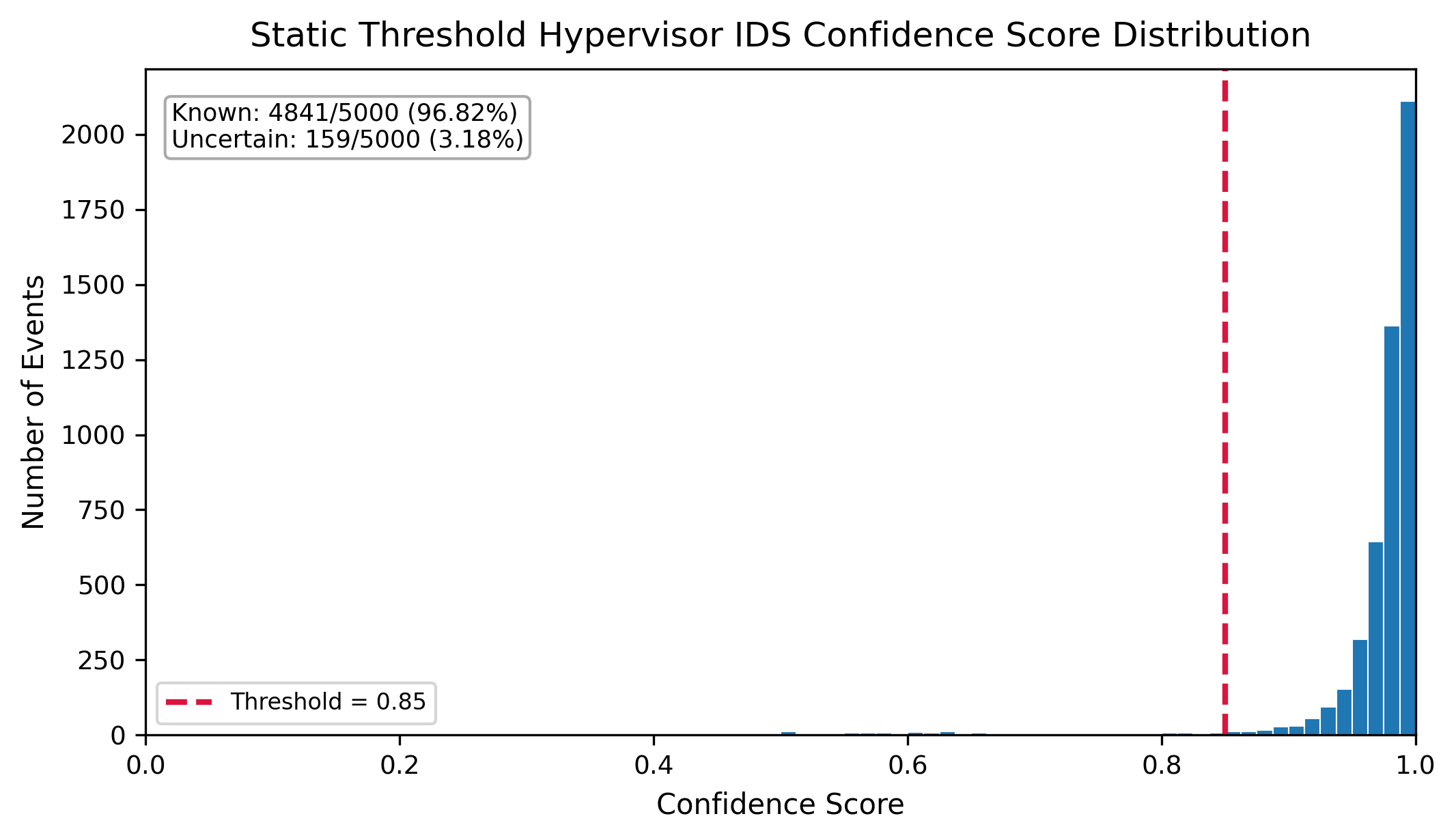}
    \caption{confidence-aware distribution at the hypervisor layer.}
    \label{fig:hypconf}
\end{figure}

Following the same 5,000-event evaluation and static-threshold protocol defined earlier for \ac{NIDS} and \ac{HIDS}, the hypervisor layer produced 4,841 known events (96.82\%) and 159 uncertain events (3.18\%), as shown in Fig~\ref{fig:hypconf}. As described earlier, events below the threshold are escalated to the \ac{LLM}; the resulting enriched outputs are then stored in ChromaDB for future retrieval and retraining support.

Table~\ref{Hyp-unc} presents representative low-confidence hypervisor events with \ac{LLM}-generated explanations and their subsequent storage in ChromaDB. The \ac{LLM} explanations presented in this work are based on the actual responses recorded in the \ac{LLM} audit logs. For readability and space considerations, they have been slightly paraphrased while preserving their original meaning and intent. The example of explanation from our results is as follows:
\begin{table}[ht]
\footnotesize
\setlength{\tabcolsep}{5pt}
\renewcommand{\arraystretch}{1.3}
\caption{Description of Hypervisor uncertain event via LLM}
\label{Hyp-unc}
\begin{tabular}{|p{2.5cm}|p{5.5cm}|}
\hline
\textbf{Field} & \textbf{Description} \\
\hline
Event Type
& Uncertain Event U2 \\
\hline
ML Confidence Score
& 0.58 \\
\hline
Routing Decision
& \textit{Uncertain} $\rightarrow$ LLM analysis \\
\hline
Model/Stage Explanation
& This event reflects ambiguous system-level behavior with partial similarity to early-stage hypervisor attack patterns, such as irregular \ac{VM} execution flow and transient low-level resource access. However, key attack characteristics such as sustained malicious activity, coordinated follow-up events, or clear exploit signatures are not observed. Consequently, the statistical model reports low confidence, and the event is escalated for semantic analysis to avoid premature or unsafe mitigation actions. \\
\hline
LLM Verdict
& Suspicious but inconclusive; manual review or contextual correlation advised. \\
\hline
\end{tabular}
\end{table}

The distribution of known and uncertain events across all three layers (\ac{NIDS}, \ac{HIDS}, and hypervisor-\ac{IDS}) at the fixed confidence threshold of 0.85 is presented in Table~\ref{tab:static-085}.
\begin{table}[ht]
\centering
\footnotesize
\setlength{\tabcolsep}{5pt}
\renewcommand{\arraystretch}{1.3}
\caption{Known and uncertain event distribution using a fixed threshold of 0.85}
\label{tab:static-085}
\begin{tabular}{|l|c|c|c|}
\hline
\textbf{Layer} & \textbf{Threshold} & \textbf{Known (\%)} & \textbf{Uncertain (\%)} \\
\hline
NIDS & 0.85 & 96.10 & 3.90 \\
\hline
HIDS & 0.85 & 53.30 & 46.70 \\
\hline
Hypervisor & 0.85 & 96.82 & 3.18 \\
\hline
\end{tabular}
\end{table}
\subsection{Q-Learning-Calibrated Proposed IDS Pipeline}
This subsection presents the complete Q-learning-calibrated \ac{IDS} pipeline, building on the framework described in Section~\ref{framework}. To address the limitations of static thresholding, an (\ac{RL})-based Q-learning approach is employed to autonomously determine layer-specific confidence thresholds. 
\subsubsection{Implementation for RL-based confidence threshold}
Following the same held-out 5,000 event evaluation protocol used for the static-threshold baseline, the \ac{RL}-based thresholding is evaluated for each \ac{IDS} layer to ensure a fair and direct comparison. The underlying \ac{XGBoost} classifier remains unchanged. The \ac{RL} calibrator operates after each \ac{XGBoost} class prediction, using the associated confidence score to decide whether the prediction is accepted as known or marked as uncertain. Table \ref{tab:threshold} shows the results for the learned threshold values and routing percentages for the proposed \ac{IDS} pipeline.

\begin{table}[ht]
\footnotesize
\setlength{\tabcolsep}{5pt}
\renewcommand{\arraystretch}{1.3}
\caption{Learned thresholds and routing percentage}
\label{tab:threshold}
\centering
\begin{tabular}{|l|c|c|c|}
\hline
\textbf{Layer} & \textbf{Threshold} & \textbf{Known (\%)} & \textbf{Uncertain (\%)} \\
\hline
NIDS & 0.81 & 96.44 & 3.56 \\
\hline
HIDS & 0.66 & 85.02 & 14.98 \\
\hline
Hypervisor & 0.87 & 96.36 & 3.64 \\
\hline
\end{tabular}
\end{table}

\begin{figure}[t]
    \includegraphics[width=\columnwidth]{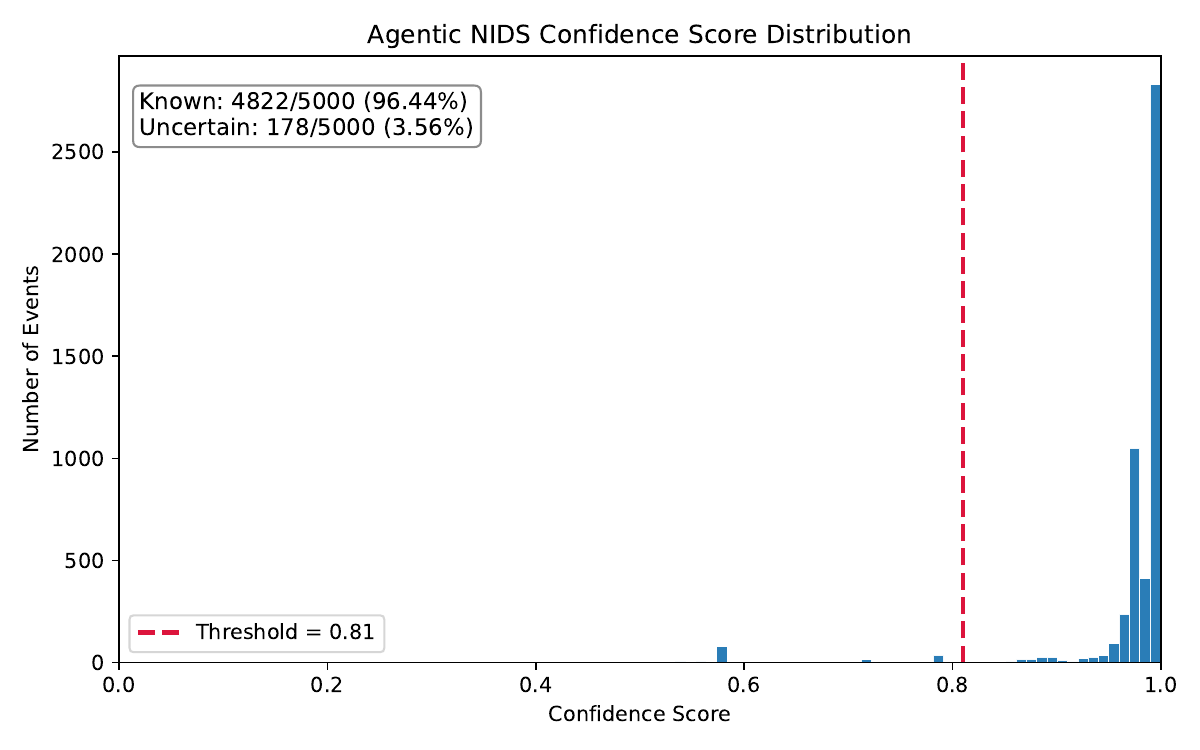}
    \caption{Distribution of prediction confidence scores produced by the Proposed NIDS component.}
    \label{fig:agentnidsconfidence}
\end{figure}
The results while implementing the \textit{proposed approach} on the network layer show that the learned threshold successfully maintained high detection while minimizing unnecessary escalation with a learned confidence threshold of 0.81, 96.44\% known events, and 3.56\% uncertain events that are forwarded to \ac{LLM}. The method achieved an average high confidence of 0.973, which depicts stable and reliable network detection. The \ac{NIDS} confidence for the events, shown in Fig~\ref{fig:agentnidsconfidence}, depicts that most of the network traffic events have a very high confidence close to 1.0, whereas only a small number of low-confidence events fall below the learned confidence threshold, which is 0.81. Furthermore, this analysis shows that the learned \ac{NIDS} threshold is operationally efficient, maintaining high-confidence acceptance while limiting uncertain escalations.

\begin{figure}[t]
    \includegraphics[width= \columnwidth]{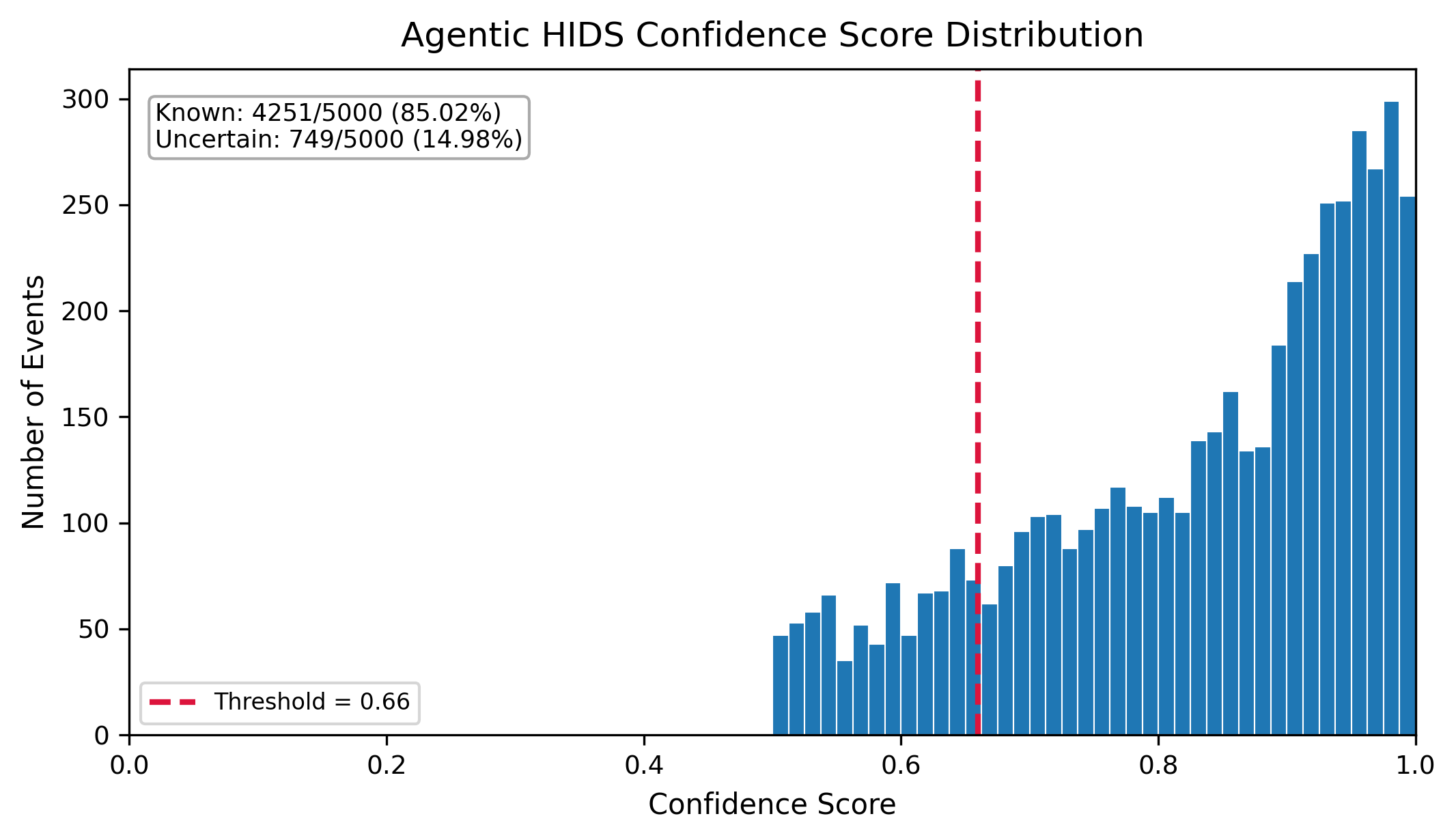}
    \caption{Distribution of prediction confidence scores produced by the Proposed HIDS component.}
    \label{fig:agenthidsconfidence}
\end{figure}
\begin{figure}[t]
    \includegraphics[width=\columnwidth]{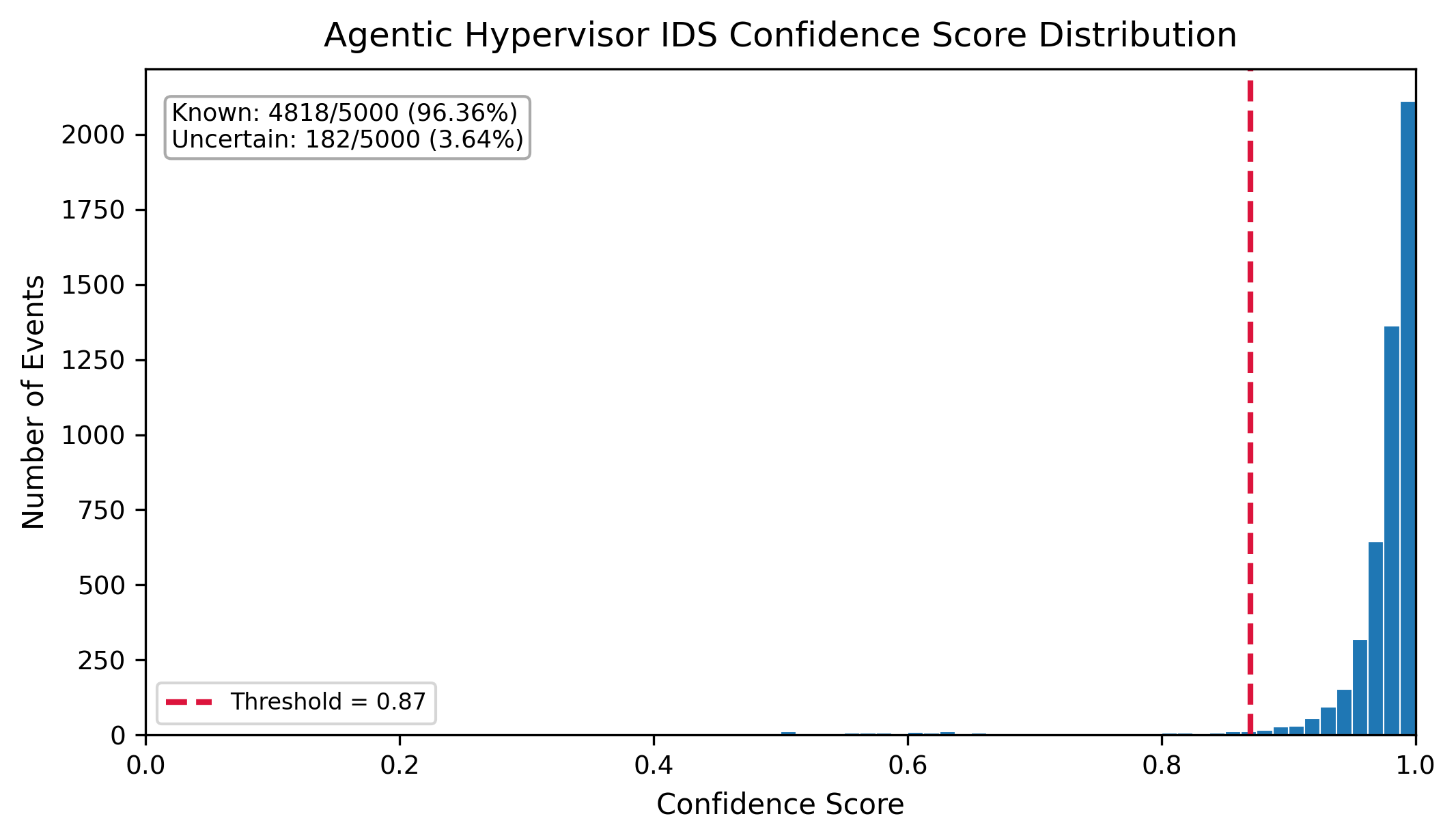}
    \caption{Distribution of prediction confidence scores produced by the Proposed Hypervisor layer component.}
    \label{fig:agenthypconfidence}
\end{figure}
\textit{For the \ac{HIDS} layer}, the learned threshold converged to 0.61, which detected 85.02\% known events and 14.98\% uncertain events. These results show that the method effectively reduced excessive \ac{LLM} escalation while preserving conservative behavior in a log-based, semantically noisy environment. The \ac{HIDS} confidence, shown in the Fig~\ref{fig:agenthidsconfidence}, demonstrates that in host-based traffic, the confidence scores are more widely spread, with notable events lying in the mid-confidence range, showing the ambiguous nature of host-based logs. The Q-learning generates the learned threshold value of 0.66, a lower adaptive threshold appropriate for log-based data, reducing unnecessary escalation to the \ac{LLM}. \textit{In the case of the hypervisor layer}, the learned threshold converged to 0.87 and generated 96.37\% events as known and 3.64\% as uncertain. This shows the effectiveness of the learned threshold to trust high-confidence hypervisor events and only escalate the genuine ambiguous cases to \ac{LLM}. The distribution of prediction confidence scores produced by the \ac{RL} based pipeline is presented in Fig~\ref{fig:agenthypconfidence}, which shows that the learned threshold clearly separates the small set of uncertain events from the large majority of high-confidence detection events. This enables largely autonomous hypervisor-level intrusion detection with minimal escalation.

\begin{figure}[t]
    \includegraphics[width=\columnwidth]{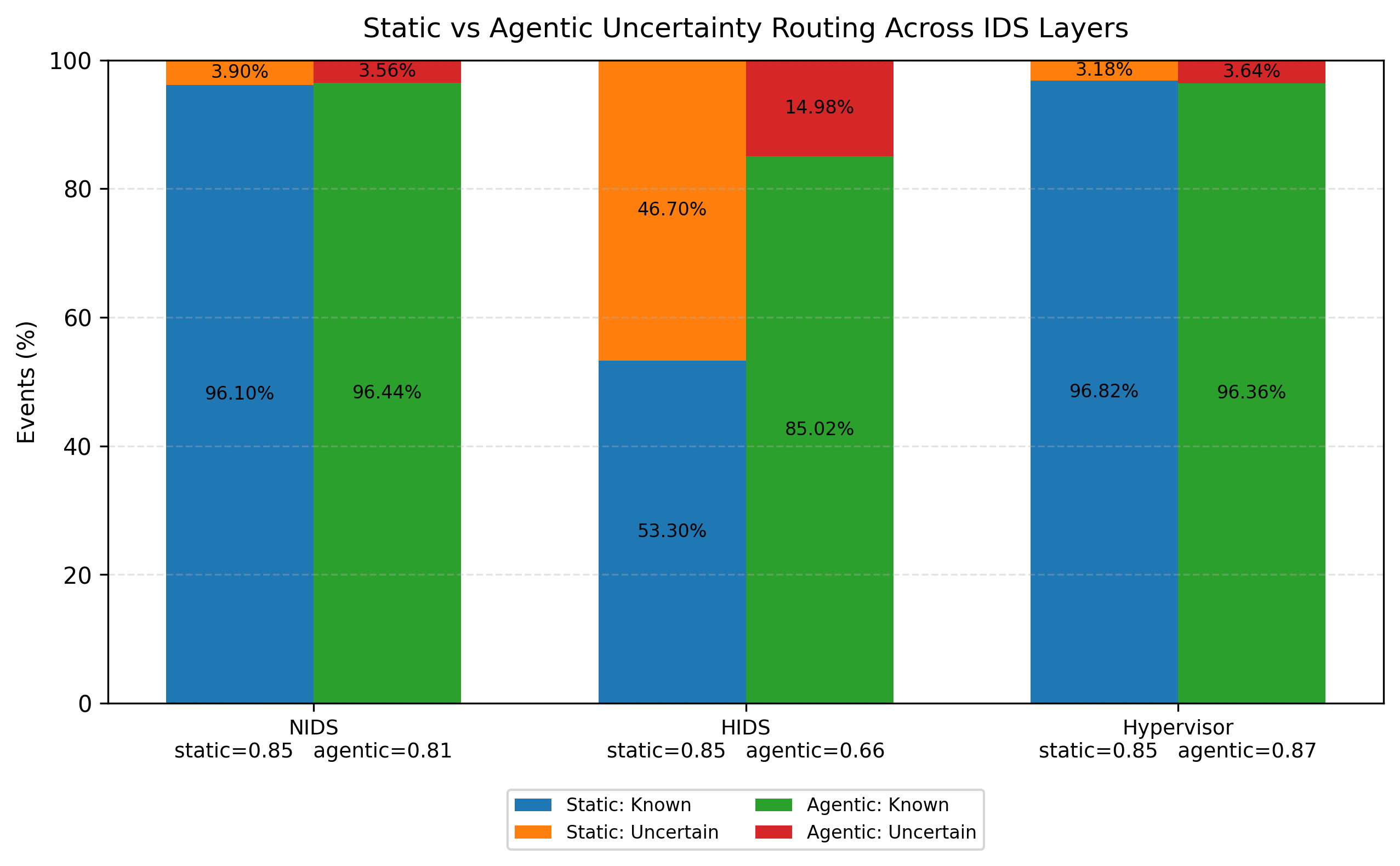}
    \caption{Baseline and proposed \ac{RL} method uncertainty routing across all layers}
    \label{fig:staticvsagentic}
\end{figure}

Comparing the proposed \ac{RL}-based pipeline with the manual pipeline, we found better results as the proposed method is managing the confidence threshold in an effective way. Table~\ref{tab:staticvsagent} shows the confidence thresholds for static and \ac{RL}-based pipeline implementations. The comparison of both baseline and proposed approaches can be visualized from Fig~\ref{fig:staticvsagentic}. This comparison clearly shows the severe over-escalation, especially in \ac{HIDS} and Hypervisor layers, due to their semantic ambiguity and non-uniform confidence distribution. \textit{The primary limitation of static confidence thresholds is the heterogeneous nature of the datasets, as all three datasets exhibit different behaviour, such as data structure, semantic ambiguity, and confidence score distributions. In contrast, the proposed \ac{RL} method works well because it can better analyze and learn the nature of datasets and can adjust the learned threshold for a balanced, resource-efficient \ac{IDS} system while maintaining high accuracy.}

\begin{table}[t]
\caption{Static vs Proposed Uncertainty Routing Across IDS Layers}
\label{tab:staticvsagent}
\centering
\footnotesize
\setlength{\tabcolsep}{5pt}
\renewcommand{\arraystretch}{1.3}
\begin{tabular}{|l|l|c|c|c|}
\hline
\textbf{Layer} & \textbf{Mode} & \textbf{Known (\%)} & \textbf{Uncertain (\%)} & \textbf{Threshold} \\
\hline
\multirow{2}{*}{NIDS}
 & Static  & 96.10 & 3.90 & 0.85 \\
\cline{2-5}
 & Proposed & 96.44 & 3.56 & 0.81 \\
\hline
\multirow{2}{*}{HIDS}
 & Static  & 53.30 & 46.70 & 0.85 \\
\cline{2-5}
 & Proposed & 85.02 & 14.98 & 0.66 \\
\hline
\multirow{2}{*}{Hypervisor}
 & Static  & 96.82 & 3.18 & 0.85 \\
\cline{2-5}
 & Proposed & 96.36 & 3.64 & 0.87 \\
\hline
\end{tabular}
\end{table}

\subsection{Events during Gate 1 and Gate 2}
After learning the thresholds successfully, if an event is detected with a confidence score higher than the learned threshold for its layer (0.81 for \ac{NIDS}, 0.66 for \ac{HIDS}, and 0.87 for the Hypervisor layer), it is tagged by the \ac{XGBoost} model as classified (known), and from there, the details for the events are stored in the output files. However, if the confidence score is lower than the learned threshold, the event is escalated to Gate 2. Gate 2 serves as the Chroma memory stage, where low-confidence events from Gate 1 are first checked against previously stored attack embeddings. If a sufficiently similar match is found, the event is tagged as a memory-supported attack match and routed without immediate \ac{LLM} escalation.
\subsection{Events at Gate 3}
Gate-3 implementation follows the calibrated \ac{LLM} validation and weighted-fusion mechanism described in Section~\ref{framework}. In the proposed pipeline, the \ac{LLM} is invoked only for unresolved low-confidence events that fail Gate-1 scrutiny and also do not satisfy the Gate-2 Chroma memory-match criteria. For each such event, a layer-aware textual representation is generated and sent to the \ac{LLM}, which returns a structured JSON response containing a semantic decision (ATTACK, BENIGN, UNSURE), a confidence score, an attack type, and a short explanation. The \ac{LLM} output is normalized and logged in an uncertain-event audit file. The layer-specific calibrated \ac{LLM} thresholds obtained through precision-constrained calibration are 0.69 for the network layer, 0.61 for the host layer, and 0.89 for the hypervisor layer. Events satisfying either the direct \ac{LLM} condition or the weighted-fusion fallback are stored as confirmed attacks in ChromaDB. At the same time, non-attack or low-confidence outcomes are routed to a low-confidence review bucket for analyst inspection.

The routing distribution obtained from the fresh run using the current thresholds is summarized as follows. \textit{For the \ac{NIDS} layer}, $178$ events were routed as uncertain, of which the \ac{LLM} classified $175$ as ATTACK and $3$ as UNSURE. All $175$ promoted attack decisions were accepted through the direct LLM condition method, while the weighted-fusion fallback contributed $0$ additional attack promotions. \textit{For the \ac{HIDS} layer}, $749$ uncertain events were forwarded to the \ac{LLM}. Among these, the \ac{LLM} classified $385$ ATTACK, $236$ BENIGN, and $128$ UNSURE events. All $385$ promoted ATTACK events were accepted through the direct LLM condition, and the weighted-fusion fallback again contributed $0$ additional attack promotions. Finally, \textit{for the Hypervisor layer}, $182$ uncertain events were analyzed by the \ac{LLM}, yielding $1$ ATTACK and $181$ UNSURE events. The single promoted attack was accepted through the direct LLM condition, while the weighted-fusion fallback contributed $0$ additional attack promotions.

Overall, the three layers produced a total of $1109$ uncertain events. Of these, the \ac{LLM} classified $561$ events as ATTACK, $236$ as BENIGN, and $312$ as UNSURE, corresponding to $50.59\%$, $21.28\%$, and $28.13\%$ of all uncertain events, respectively. All $561$ promoted attacks were accepted through the direct LLM condition method, whereas the weighted fallback mechanism contributed $0$ additional attack decisions. The reason the weighted fallback remained inactive in this run is that it is only applied when the \ac{LLM} predicts ATTACK but with confidence below the direct layer-specific \ac{LLM} threshold. In the present run, this condition did not occur in any layer. Every \ac{LLM} output labeled as ATTACK already exceeded the direct confidence threshold. Therefore, all promoted attacks were resolved through the primary LLM direct condition without invoking weighted fusion.
\subsection{Post-Gates Process}
If an event does not satisfy the Gate 3 attack-promotion criteria, it is not discarded. Instead, it is routed to a low-confidence or non-attack review bucket and saved in JSONL format for analyst inspection. In the final run, this review bucket contained $548$ events in total, comprising $3$ events from \ac{NIDS}, $364$ from \ac{HIDS}, and $181$ from the Hypervisor layer. These events correspond to cases where the \ac{LLM} remained uncertain or where the response did not provide sufficient confidence for automatic escalation. Retaining these records supports traceability, manual validation, and later model improvement.
Finally, the run produces structured outputs including confidence CSVs, uncertain-event logs, low-confidence review files, and a run summary. These additional documents support reproducibility, analyst review, and iterative improvement. Over time, confirmed attacks added to ChromaDB and reviewed low-confidence cases can be used to expand the knowledge base and inform future retraining and recalibration of thresholds and fusion weights.

\section{Results Analysis}
\label{res}
 In our study, we have found that the proposed \ac{RL}-based pipeline is considered more cost-efficient because it reduces unnecessary escalations to the \ac{LLM} while maintaining acceptable detection performance.

\noindent\textbf{Observed Escalation Analysis:}
For the experimental setup with 5,000 test events per layer, the proposed \ac{RL}-based routing with adaptive thresholds produced escalation counts of 178 (\ac{NIDS}), 749 (\ac{HIDS}), and 182 (Hypervisor), whereas static routing at a 0.85 threshold produced 195 (\ac{NIDS}), 2335 (\ac{HIDS}), and 159 (Hypervisor).
Thus, the total number of uncertain events escalated by the proposed scheme is,
\begin{equation}
N_{\mathrm{unc}}^{\mathrm{RL}} = 178 + 749 + 182 = 1109,
\end{equation}
and for the static scheme is, 
\begin{equation}
N_{\mathrm{unc}}^{\mathrm{static}} = 195 + 2335 + 159 = 2689.
\end{equation}
The reduction in escalations achieved by the proposed \ac{RL}-based pipeline can be simply given as,
\begin{equation}
\Delta N_{\mathrm{unc}} = N_{\mathrm{unc}}^{\mathrm{static}} - N_{\mathrm{unc}}^{\mathrm{RL}} =2689 - 1109 = 1580,
\end{equation}
and the percentage reduction is therefore,
\begin{equation}
\text{Reduction (\%)} = \frac{\Delta N_{\mathrm{unc}}}{N_{\mathrm{unc}}^{\mathrm{static}}}\times 100=\frac{1580}{2689} \times 100 = 58.76\%.
\end{equation}
Across 15,000 evaluated events, the proposed pipeline routing cuts escalations from 2,689 to 1,109, reducing expensive \ac{LLM}/analyst handoffs by 58.76\%. This lowers escalation intensity from 17.9\% to 7.4\% of all events.
 
\noindent\textbf{Estimated Cost Reduction Analysis:} Since \ac{LLM} inference APIs are billed per token~\cite{openai2024pricing, anthropic2024pricing}, the total inference cost $(C_{\mathrm{total}})$ scales directly with the number of escalated events. For a routing policy producing $N_{\mathrm{unc}}$ escalations, the total cost is:
\begin{equation}
C_{\mathrm{total}} = N_{\mathrm{unc}} \times c_{\mathrm{event}},
\label{eq:total_cost}
\end{equation}
where $c_{\mathrm{event}}$ is the cost per escalated event. The cost saving of the proposed \ac{RL}-based pipeline over the static scheme can be calculated as,
\begin{equation}
\Delta C = \bigl(N_{\mathrm{unc}}^{\mathrm{static}} - N_{\mathrm{unc}}^{\mathrm{RL}}\bigr) \times c_{\mathrm{event}} = 1{,}580 \times c_{\mathrm{event}}.
\label{eq:cost_saving}
\end{equation}

Since $c_{\mathrm{event}}$ is a fixed cost determined by the choice of \ac{LLM} provider and prompt design~\cite{intuitionlabs2025pricing}, eq.~(\ref{eq:cost_saving}) shows that cost saving scales linearly with escalation reduction. The proposed pipeline reduces escalations by 58.76\%, yielding a direct reduction in LLM inference expenditure, without any compromise to detection performance (Table~\ref{tab:escalation-comparison}).
\begin{table}[ht]
\centering
\footnotesize
\setlength{\tabcolsep}{5pt}
\renewcommand{\arraystretch}{1.3}
\caption{Escalation count comparison between static and RL-based routing.}
\label{tab:escalation-comparison}
\begin{tabular}{|l|c|c|}
\hline
\textbf{Mode} & \textbf{Total Escalations} & \textbf{Reduction (\%)} \\
\hline
Static & 2,689 & -- \\
\hline
Proposed & 1,109 & 58.76 \\
\hline
\end{tabular}
\end{table}

\noindent\textbf{Routing Selectivity and Detection Performance:}
Table~\ref{tab:cold_run_routing} shows that the proposed pipeline keeps escalation selective rather than indiscriminate. Only 1,109 of 15,000 events (7.39\%) are routed as uncertain, while 13,891 events are handled directly by the base detectors. Among uncertain events, 561 are promoted by \ac{LLM} analysis, and 548 are retained in a low-confidence bucket for review, which avoids forcing unreliable automatic decisions. This behavior is important for operational \ac{IDS} settings because it balances automation with risk control.
\begin{table}[h]
\centering
\footnotesize
\setlength{\tabcolsep}{3pt}
\renewcommand{\arraystretch}{1.3}
\caption{Routing and escalation outcomes across the three IDS layers.}
\label{tab:cold_run_routing}
\begin{tabular}{|C{1cm}|C{1cm}|C{1.1cm}|C{1.1cm}|C{1cm}|C{1.1cm}|C{0.9cm}|}
\hline
\textbf{Layer} &
\makecell{\textbf{Total}\\\textbf{events}} &
\makecell{\textbf{Uncert.}\\\textbf{events}} &
\makecell{\textbf{Uncert.}\\\textbf{(\%)}} &
\textbf{LLM-detected} &
\makecell{\textbf{Low-conf.}\\\textbf{bucket}} &
\makecell{\textbf{Bucket}\\\textbf{(\%)}} \\
\hline
NIDS & 5,000 & 178 & 3.56 & 175 & 3 & 0.06 \\
\hline
HIDS & 5,000 & 749 & 14.98 & 385 & 364 & 7.28 \\
\hline
HypIDS & 5,000 & 182 & 3.64 & 1 & 181 & 3.62 \\
\hline
Overall & 15,000 & 1,109 & 7.39 & 561 & 548 & 3.65 \\
\hline
\end{tabular}
\end{table}

Table~\ref{tab:cold_run_performance} presents the strong end-to-end detection quality of the proposed routing policy with an overall accuracy of 88.68\%, precision of 85.29\%, recall of 84.72\%, and F1 score of 85.00\%. At the layer level, \ac{NIDS} and Hypervisor-based IDS record excellent F1 score with  96.51\% and 96.97\%, respectively. However, \ac{HIDS} F1 score remains low (60.40\%) due to noisier host-level behavior and higher class ambiguity. Importantly, instead of over-committing on difficult \ac{HIDS} cases, the framework adaptively increases uncertainty routing at \ac{HIDS} (14.98\%) compared to \ac{NIDS} (3.56\%) and Hypervisor-based IDS (3.64\%), which is the intended risk-aware behavior.
\begin{table}[ht]
\centering
\footnotesize
\setlength{\tabcolsep}{4pt}
\renewcommand{\arraystretch}{1.3}
\caption{End-to-End detection performance of the proposed three-layer  IDS pipeline.}
\label{tab:cold_run_performance}
\begin{tabular}{|C{1.2cm}|C{1.2cm}|C{1.2cm}|C{1.2cm}|C{1.2cm}|}
\hline
\textbf{Layer} & \textbf{Acc. (\%)} & \textbf{Prec. (\%)} & \textbf{Rec. (\%)} & \textbf{F1 (\%)} \\
\hline
NIDS & 98.02 & 97.43 & 95.59 & 96.51 \\
\hline
HIDS & 70.94 & 58.81 & 62.07 & 60.40 \\
\hline
HypIDS & 97.08 & 99.24 & 94.81 & 96.97 \\
\hline
Overall & 88.68 & 85.29 & 84.72 & 85.00 \\
\hline
\end{tabular}
\end{table} 

\noindent\textbf{Cross-layer adaptation statement:}
The proposed method is evaluated across three heterogeneous environments and datasets: network-flow data (CICIDS2018), host-log data (LIDDS2019), and synthetic hypervisor events with injected boundary/noise effects. Despite different feature spaces, attack semantics, and noise levels, the same three-stage framework adapts per layer and maintains stable overall performance. This demonstrates that the method is not tuned to a single dataset type; it generalizes across network, host, and virtualisation monitoring contexts by using layer-specific learned thresholds and uncertainty-aware escalation.
\subsection{Comparison with Existing Studies}

Drawing on the literature review in Section \ref{sec:Lit}, relevant \ac{LLM}-assisted intrusion detection studies were identified and compared with the proposed pipeline. The analysis indicates limited overlap, with notable differences in system scope and architectural design. \citep{FernandezSaura2026FGCS} escalates low-confidence alerts to a local \ac{LLM} within a federated setting, but operates on a single NIDS layer with a fixed confidence threshold of 0.85, and does not quantify escalation cost. \citep{Kalafatidis2025} applies a two-tier LLM-enhanced detection strategy for containerised SDN  environments, triggering on-demand \ac{LLM} analysis via a static statistical threshold, with no host or hypervisor coverage and no adaptive routing. \citep{adjewa2025llm} dynamically identifies unknown attack clusters using transformer encoders and Gaussian mixture models, achieving adaptive identification without retraining, but remains confined to network-level detection with no \ac{LLM} escalation control or threshold learning. \citep{Tavallaee2024HybridNIDSHIDS} combines \ac{NIDS} and \ac{HIDS} through a two-stage collaborative classifier with NLP-based host data transformation, extending coverage to two layers, yet still relies on a fixed decision boundary with no \ac{RL}-based optimisation or \ac{LLM} escalation gate. The details of the comparison of the proposed study with the latest existing studies are shown in Table~\ref{tab:comparison}.
\begin{table*}[!t]
\footnotesize
\setlength{\tabcolsep}{4pt}
\renewcommand{\arraystretch}{1.25}
\caption{Comparison of the proposed pipeline with related works.}
\label{tab:comparison}
\centering
\begin{tabularx}{\textwidth}{|>{\raggedright\arraybackslash}X|c|c|c|c|c|}
\hline
\textbf{Criterion} &
\makecell{\textbf{Fern{\'a}ndez Saura}\\\citep{FernandezSaura2026FGCS}} &
\makecell{\textbf{Kalafatidis}\\\citep{Kalafatidis2025}} &
\makecell{\textbf{Adjewa}\\\citep{adjewa2025llm}} &
\makecell{\textbf{Tavallaee}\\\citep{Tavallaee2024HybridNIDSHIDS}} &
\makecell{\textbf{Proposed}\\\textbf{Work}} \\
\hline
Three-layer IDS (NIDS+HIDS+HypIDS) & \xmark & \xmark & \xmark & \xmark & \cmark \\
\hline
RL-based adaptive threshold per layer & \xmark & \xmark & \xmark & \xmark & \cmark \\
\hline
Selective LLM escalation & \cmark & \cmark & \xmark & \xmark & \cmark \\
\hline
Borderline cases decision mechanism & \xmark & \xmark & \xmark & \xmark & \cmark \\
\hline
Fixed-threshold independence & \xmark & \xmark & \xmark & \xmark & \cmark \\
\hline
Explicit escalation-cost analysis & \xmark & \xmark & \xmark & \xmark & \cmark \\
\hline
Static-threshold baseline reported & \xmark & \xmark & \xmark & \xmark & \cmark \\
\hline
Hypervisor layer coverage & $\sim$ & \cmark & \xmark & \xmark & \cmark \\
\hline
\end{tabularx}
\vspace{1pt}
\raggedright\footnotesize{$\sim$ = partially addressed,\;
\cmark = present,\;
\xmark = absent.}
\end{table*}

The proposed work addresses the limitations across all four baselines simultaneously. To the best of our knowledge, it is the only pipeline that covers three \ac{IDS} layers, applies Q-learning to independently learn confidence thresholds per layer, routes uncertain events selectively to an \ac{LLM} without any fixed threshold, retains unresolved low-confidence events in a bucket, and reports explicit escalation-cost efficiency against a static baseline. This combination of adaptive multi-layer routing and cost-aware \ac{LLM} escalation was not comprehensively reported in prior works.
\section{Conclusions}
\label{con}
This study presented a cloud-specific multi-layer IDS integrating XGBoost and LLM reasoning within a confidence-aware multi-gated architecture. Different from the existing approaches that rely on fixed classifiers and static LLM thresholds, the proposed framework employs Q-learning for adaptive threshold calibration per-layer, and weighted-fusion fallback for borderline cases, enabling more reliable decision control across network, host, and hypervisor layers. Experimental results showed that the proposed pipeline reduced unnecessary LLM escalations by 58.76\% while maintaining strong detection performance, achieving 88.68\% overall accuracy and up to 98.02\% accuracy at the network layer. The framework further improved operational efficiency by cautiously promoting uncertain attacks and routing unresolved low-confidence events to analyst review rather than forcing risky automatic decisions. These results demonstrate the practical value of combining RL, confidence calibration, and LLM-assistance for cloud intrusion detection. Future work will focus on adversarial robustness, zero-day attack resilience, and real-time deployment in dynamic cloud environments.
\printglossaries
\bibliographystyle{ieeetr}  
\bibliography{references}

\end{document}